\documentclass[11pt,a4paper,onecolumn]{article}

\usepackage[version=3]{mhchem} 
\usepackage[T1]{fontenc}       
\usepackage{graphicx}
\usepackage{dblfloatfix}
\usepackage{placeins}
\usepackage{authblk}
\usepackage{a4wide}
\usepackage{setspace}
\usepackage[margin=1in]{geometry}

\author[1]{Vivek Tiwari}
\author[2]{Yassel Acosta Matutes}
\author[3]{Alastair T. Gardiner}
\author[3]{Richard J. Cogdell}
\author[1]{Jennifer P. Ogilvie\thanks{jogilvie@umich.edu}}
\affil[1]{Department of Physics, University of Michigan, Ann Arbor, Michigan, 48109, United States}
\affil[2]{Applied Physics Program, University of Michigan, Ann Arbor, Michigan, 48109, United States}
\affil[3]{Institute for Molecular Biology, University of Glasgow, Glasgow G12 8TA, United Kingdom}

\begin{document}

\title {Spatially-resolved Fluorescence-detected Two-dimensional Electronic Spectroscopy Probes Varying Electronic Couplings in Photosynthetic Bacteria }
\maketitle

\doublespacing

\begin{abstract}
 We present a variation of two-dimensional electronic spectroscopy that is capable of mapping spatially-varying differences in electronic couplings using a correlated map of excitation and detection frequencies, with sensitivity orders of magnitude better than conventional spatially-averaged electronic spectroscopies. The approach performs fluorescence-detection-based fully collinear two-dimensional electronic spectroscopy in a microscope, combining femtosecond time-resolution, sub-micron spatial resolution, and the sensitivity of fluorescence detection. We demonstrate the approach on a mixture of photosynthetic bacteria that are known to exhibit variations in electronic structure with growth conditions. Spatial variations in the constitution of mixed bacterial colonies manifests as spatially-varying peak intensities in the measured two-dimensional contour maps, which exhibit well-resolved electronic couplings between excited electronic states of the bacterial proteins. \end{abstract}

\section*{Introduction}
Ultrafast spectroscopy \cite{Zewail1985,Mukamel_book}, has contributed to the fundamental understanding of a wide range of processes including the primary steps of vision \cite{Mathies1991}, energy transfer and charge separation in natural \cite{Blankenship_book, Photosynthetic_excitons_book} and artificial light-harvesting systems, and carrier relaxation pathways in semiconductors \cite{Woggon1996,Spencer2008,Zhang2002}. Understanding the ultrafast electronic dynamics between excited electronic states of molecules is important in many fields, ranging from artificial \cite{Silva2014,Friend2017,Miller2017} and natural light harvesting \cite{Scholes2017} to the chemistry of melanin pigments at cancerous tumor sites \cite{Warren2011}.  Although pump-probe spectroscopies can capture ultrafast relaxation processes, important information such as broadening mechanisms and couplings between excited electronic states that underlie ultrafast processes are obscured in such measurements\cite{Jonas2003,OgilvieKubarych2009}. 

The invention of two-dimensional electronic spectroscopy (2DES) \cite{Jonas2003}, which is an optical analog \cite{JonasScience2003} of 2D nuclear magnetic resonance (NMR) \cite{Ernst1976}, has provided high experimental frequency resolution sufficient to distinguish many similar transient chemical species based on their correlated absorption and subsequent emission properties. By correlating the initial absorption frequencies of a system with subsequent changes along a detection frequency axis, a 2D contour map of highly-resolved excitation and detection frequency axes can decongest the couplings that cause transitions between electronic states, to the extent that is limited only by the molecular resolution. Due to the added dimensionality, the information content of a 2DES experiment is a superset of what is available from pump-probe spectroscopy. 2DES experiments have been broadly applied, and have contributed to the mechanistic understanding of delocalized electronic  and mixed electronic-vibrational states in natural photosynthetic systems  \cite{Engel2007,Fuller2014,Womick2011,Dean2017,Palecek2017}, molecular aggregates \cite{Hauer2015}, hybrid organic/inorganic perovskites \cite{Friend2017,Miller2017} and singlet fission materials \cite{Rao2016}.

Despite its success in advancing fundamental understanding of a variety of condensed phase phenomena, most 2DES experiments have offered no spatial resolution (typically a few hundreds of microns). This limitation produces 2DES signals that represent an ensemble-averaged response over a large number of species. Many of the condensed phase systems for which 2DES experiments can provide the deepest insight possess highly disordered spatial and/or energetic landscapes, rendering many experimental conclusions susceptible to ensemble averaging effects. For instance, in photosynthetic systems, where the protein environment imparts disorder in pigment electronic excited state energies, coherent dynamics between such states will `artificially' damp because the measured 2DES response will be a weighted average over the entire disordered energetic landscape. Such `ensemble dephasing' has been experimentally observed between a pair of excitons in a photosynthetic protein \cite{Savikhin1997, ThyrhaugFMO2017}. A time-resolved molecular level understanding of the role of morphology on the performance of a number of artificial photovoltaic materials is also limited due to spatial averaging. For instance, heterogeneous domain composition exists in polymer/fullerene solar cells and modulates exciton dissociation and charge recombination on picosecond timescales \cite{Asbury2014, Rao2015}. Similarly, the effect of grain size \cite{Mohite2015} and chloride content \cite{Snaith2014} on charge transport in perovskites thin-films is well known, but the connections between layer heterogeneity and charge delocalization is poorly understood. Spatial-resolution provided by pump-probe microscopy has partly addressed these questions \cite{Wong2015, Huang2015, Kukura2016, Warren2011}. However, new spatially-resolved tools that can go beyond pump-probe approaches to resolve inhomogeneity and the couplings between excited electronic states directly with high frequency and time-resolution could impact our understanding of a broad range of materials.

Here, we present spatially-resolved fluorescence-detected two-dimensional electronic spectroscopy (SF--2DES) which combines the femtosecond time resolution of a broadband laser pulse and the frequency resolution of 2D spectroscopy with spatial resolution beyond that of a two-photon microscope. The use of rapid phase-modulation and lock-in detection enables the use of high repetition rate lasers compatible with imaging applications and produces high signal-to-noise ratio images. We demonstrate \emph{in vivo} measurements on a mixture of photosynthetic bacterial cells from \emph{Rps. palustris} grown under different light intensity conditions. SF-2DES reveals spatially-varying level couplings (that manifest as well-resolved 2D peaks) as a function of spatial heterogeneity in the constitution of the bacterial colony. By adopting a fluorescence-detection approach, as opposed to radiated electric-field detection in conventional space-averaged 2DES measurements, we estimate nearly six orders of magnitude fewer bacterial cells contribute to the signal compared to conventional 2D measurements on similar systems \cite{Zigmantas2016}. We show that our measurements can differentiate between the growth condition dependent perturbations in the excitonic structure of constituent bacteria with a high signal-to-noise ratio.

\section*{Results}
\subsection*{Methodology} 

A 2D experiment \cite{OgilvieARPC} is comprised of three pulses with precisely controlled time-spacing between the pulses. Each pulse interacts with the sample up to $1^{st}$ order in time-dependent perturbation theory. When there is a manifold of coupled excited electronic states, the effect of the first two `pump' pulses can be understood as preparing a superposition of excited (or ground) states, which is then allowed to evolve during the time interval between the second and third pulses, typically called the pump-probe waiting time. The third pulse probes the evolving superposition by generating a $3^{rd}$ order macroscopic polarization in the sample at different pump-probe waiting times. In the presence of an ensemble of dipoles, the fields radiated by individual oscillating dipoles within the macroscopic polarization coherently add up to generate an experimentally detectable electric field signal along a specific phase-matched direction, which depends on a linear combination of the wavevectors of the individual pulses, that is, $\hat{k}_{sig,R} = -\hat{k}_1+\hat{k}_2+\hat{k}_3$ and $\hat{k}_{sig,NR} = +\hat{k}_1-\hat{k}_2+\hat{k}_3$, where $\hat{k}_{sig,R}$ and $\hat{k}_{sig,NR}$ are rephasing and non-rephasing 2D signals, and $\hat{k}_i, i=1-3$ are wavevectors of the three pulses. Phase-matching allows background free detection, providing a route towards high signal-to-noise ratios. However, phase-matching also constrains the 2D experiment. The generation of a phase-matched signal intrinsically relies on probing dipole ensembles with volumes larger than $\lambda^3$, where $\lambda$ is the wavelength of light. In addition, background free detection through phase-matching requires that the pulses be non-collinear, making the experiment difficult to integrate with a microscope objective, thus constraining the possibility of spatially-resolved measurements.

Fluorescence detected two-dimensional electronic spectroscopy (F--2DES) in a fully collinear geometry was first implemented by Warren and co-workers using a phase-cycled acousto-optic modulator (AOM) based pulse-shaping approach \cite{Wagner2005}. Fluorescence detection implies that sample volumes smaller than $\lambda^3$ can be detected as the need for a macroscopic grating of dipoles radiating a detectable electric field is replaced by the detection of fluorescence from an excited state population that is produced by the addition of a $4^{th}$ pulse. A fully collinear geometry implies that the setup can be integrated with a microscope objective, making spatially resolved measurements facile. In the fully collinear geometry, background-free detection can be achieved by replacing phase-matching with phase-cycling \cite{Wagner2005} or phase-modulation \cite{Tekavec2007}. Several phase-cycling approaches to F--2DES have been demonstrated \cite{Wagner2005,Brixner2017,De2014}.
To date, phase-cycling methods have been employed in spatially-resolved 2D-infrared vibrational spectroscopy in the ground electronic state of a metal-carbonyl stained polystyrene bead \cite{Zanni2016,Baiz2014}. Very recently, Goetz et al. \cite{Brixner2018} performed phase-cycling-based F--2DES in a microscope.  Some of the phase-cycling-based approaches that have been used for imaging applications have relied on a single AOM pulse-shaper to create multiple time-delayed pulses.   Depending on the pulse-shaper, this may constrain the method to only a few tens of kHz laser repetition rates. In addition, control over the polarization and spectrum of each pulse, both of which have been routinely exploited to suppress unwanted signals \cite{OgilvieARPC}, may become difficult to implement. While better for use with high repetition rate lasers, spatial-light-modulator pulse-shapers have relatively slow switching times, making setups that employ them susceptible to laser noise. In addition, phase-cycling approaches often require that as many as 27 scans with different relative pulse phases be acquired to extract a single absorptive 2D spectrum \cite{Brixner2017,Brixner2018}. For samples where photobleaching is significant, this requirement is particularly problematic.

\begin{figure*}[]
	\centering
	\includegraphics[width=6 in]{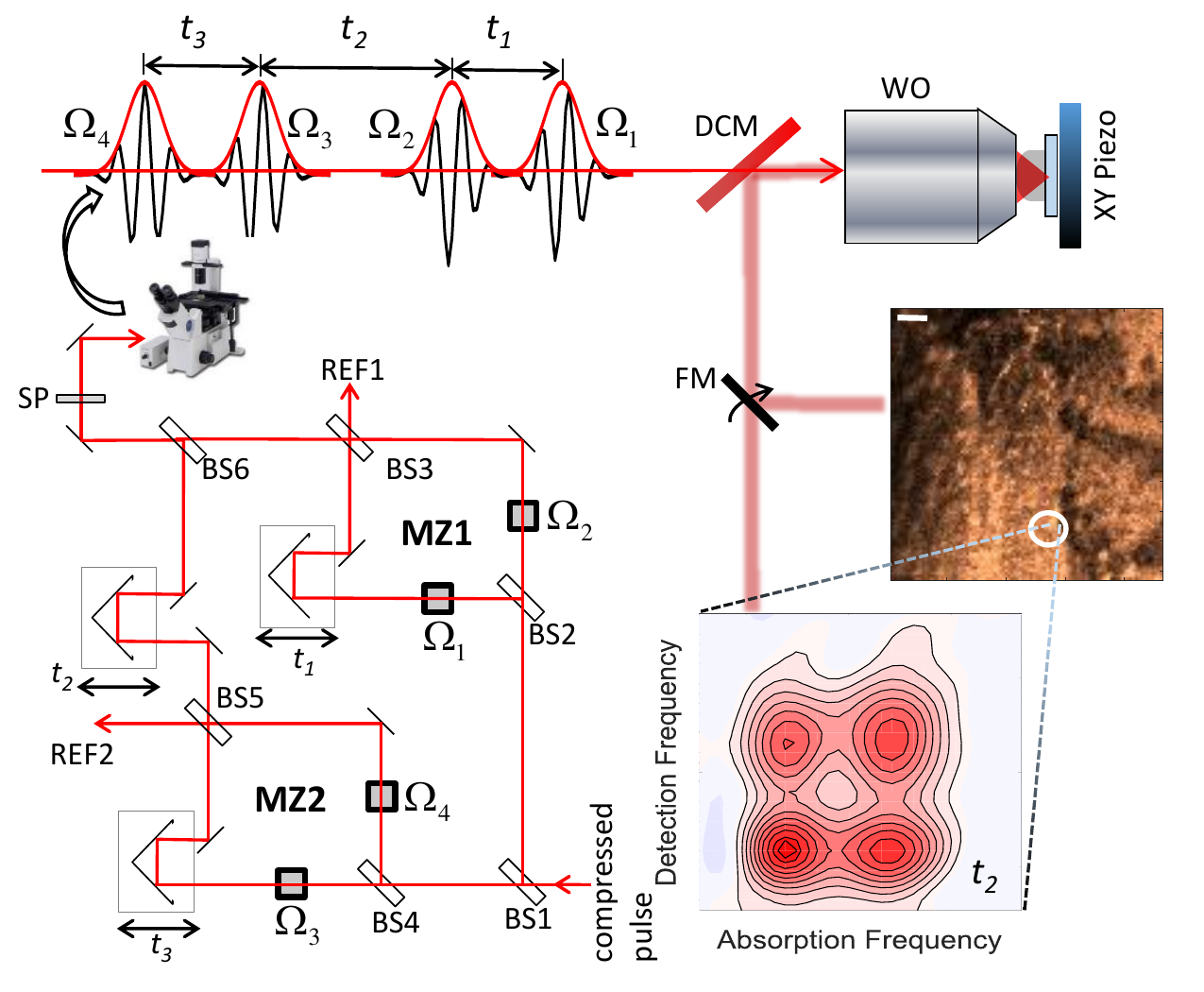}
	\caption{A simplified description of the spatially-resolved fluorescence-detected 2DES spectrometer (SF--2DES). Further details of the spectrometer are provided in the Supplementary Information (SI). A given pulse in the compressed laser pulse train is split 50:50 by a beam-splitter (BS1), and each half is routed into a Mach-Zehnder (MZ) interferometer (MZ1 and MZ2). Each of the four interferometer arms (two per MZ) contains an AOM which sweeps the carrier-envelope phase of the pulse by frequency $\Omega_i, i=1-4$. The time intervals between the four pulses, $t_1$, $t_2$ and $t_3$ are controlled by mechanical delay stages. One output port from each MZ is used to generate a reference signal REF1(2) which is utilized by the lock-in amplifier for signal detection. The other output port from each MZ is combined at BS6, generating four collinear time separated pulses (pump and probe pulse pairs), which are optically filtered by a shortpass filter (SP), and routed into a confocal microscope. A dichroic mirror (DCM) in the microscope transmits the collinear pulse train towards a water objective (WO), which focuses it on an immobilized sample. The sample is mounted on an XY scanning piezo stage (PZ). The fluorescence collected by the WO is separated from the excitation light at the DCM, and can be either routed for fluorescence imaging, or for generating a 2D map. An example of the fluorescence image, and the 2D spectrum at a desired XY location is shown in the figure. The 2D spectrum corresponds to zero waiting time between pump and probe pulse pairs ($t_2=0$), and shows absorptive changes in the refractive index of the sample in the form of well-resolved 2D peakshapes. Cross peaks at $t_2=0$ indicate that the absorption and detection frequencies of the system are different. This implies that the transitions corresponding to the positions of the two diagonal peaks correspond to excitonic transitions between sites which are electronically coupled on the excited state, and therefore connected via a common ground electronic state, and a common doubly-excited electronic state manifold. 
	}
	\label{fig:fig1}
\end{figure*}
 
Instead of phase-cycling, we adopt the alternative phase-modulation approach to F-2DES demonstrated by Marcus and co-workers \cite{Tekavec2007} for our implementation of SF-2DES. A simplified layout of the setup is shown in Fig.\ref{fig:fig1}. Four time-delayed pulses are created using interferometers, MZ1 and MZ2, and the carrier-envelope phase of each pulse, $\Omega_i, i=1-4$ is scanned by its respective AOM, over the laser pulse train. Thus, each pulse is tagged with a unique radio-frequency $\Omega_i$, which replaces the unique wavevectors, $\hat{k}_i$ of a 2DES experiment. The resulting rephasing and non-rephasing 2D signals are contained in a four-wave mixing (FWM) population which modulates at the linear combination of radio frequencies of the individual pulses, that is, $\Omega_R = -\Omega_1 + \Omega_2 + \Omega_3 -\Omega_4$ and $\Omega_{NR} = \Omega_1 - \Omega_2 + \Omega_3 -\Omega_4$, for rephasing and non-rephasing signals respectively. The two signals are demodulated and detected in parallel lock-in channels using phase-sensitive lock-in detection \cite{Blair1975}. The signal is physically undersampled through detection relative to a reference wavelength, generated from REF1 and REF2 outputs of MZ1 and MZ2 respectively (see Methods, SI for details). The reference frequencies are chosen close to the electronic energy gap, such that the signal phase oscillates at $(\omega_{eg}-\omega_{R1(2)})$, where $\omega_{eg}$ is the electronic energy gap which is sampled during time delay $t_{1(3)}$, and $\omega_{R1(2)}$ are reference frequencies generated using REF1(2). Undersampling makes the measurement insensitive to phase noise caused by mechanical delay fluctuations in the interferometer arms, thus avoiding the need for active-phase stabilization \cite{Cundiff2005}. In contrast to the approaches mentioned above, there is no constraint of kHz pulse repetition rates, and the polarization and spectrum of each pulse can be independently controlled with ease. Use of a lock-in amplifier for phase-sensitive signal detection is another major advantage that allows high signal detection sensitivity over a wide dynamic range. Modulation of the resulting signal at high frequencies implies that $1/f$ noise can be minimized, and frequency filtering by lock-in detection enables minimization of white noise. Moreover, phasing of the 2D signal is done in the time-domain at $t_1, t_2, t_3 = 0$ by the lock-in amplifier, making it substantially easier than many other approaches to 2DES \cite{OgilvieARPC}. Importantly, in the phase-modulation approach \cite{Tekavec2007}, the phase-cycling happens in `real time' as each AOM sweeps the carrier-envelope phase over the laser pulse train. This obviates the need in the phase-cycling approach \cite{Wagner2005,Brixner2017,De2014,Brixner2018} for combining multiple separate phase scans during which time photobleaching and laser fluctuations will reduce the signal-to-noise ratio with which the desired signal can be extracted. 

To perform SF-2DES, the collinear phase-modulated pulse train generated by MZ1 and MZ2 is routed to a confocal microscope as shown in Fig. \ref{fig:fig1}. First a fluorescence map of the sample is generated, followed by acquisition of F-2DES spectra at the desired XY locations. 
\FloatBarrier
\subsection*{SF--2DES on unmixed samples}

Chromatic adaptation in a number of species of purple photosynthetic bacteria involves both a growth in the size of the photosynthetic unit through an increase in the number of peripheral light-harvesting antenna complexes (LH2) per reaction center core (RC-LH1 complex) \cite{Scheuring2005}, as well as synthesis of LH2 complexes with modified spectral properties \cite{Cogdell2009}. Under high-light (HL) conditions the LH2 complex contains a monomeric ring of 9 bacteriochlorophyll a (\emph{BChl a}) pigments which absorbs at $\sim$800 nm (B800 ring), and a dimeric ring of 18 \emph{BChl a} pigments which absorbs at $\sim$850 nm (B850 ring) at 300 K. The pigments are held together by nine $\alpha\beta$ polypeptide pairs whose composition is light intensity dependent, and is dictated by which of the multigene family of \emph{puc} genes are expressed \cite{Southall2018}.  \emph{Rps. palustris} presents an interesting case because under lower light (LL) intensity conditions, the B850 ring loses oscillator strength and a band around 810 nm appears. This has been attributed to the polypeptide inhomogeneity within an LH2 ring, causing a blue-shift in the site energies of certain \emph{BChl a} pigments \cite{CogdellSM2009}.  

\begin{figure*}[h!]
	\centering
	\includegraphics[width=6 in]{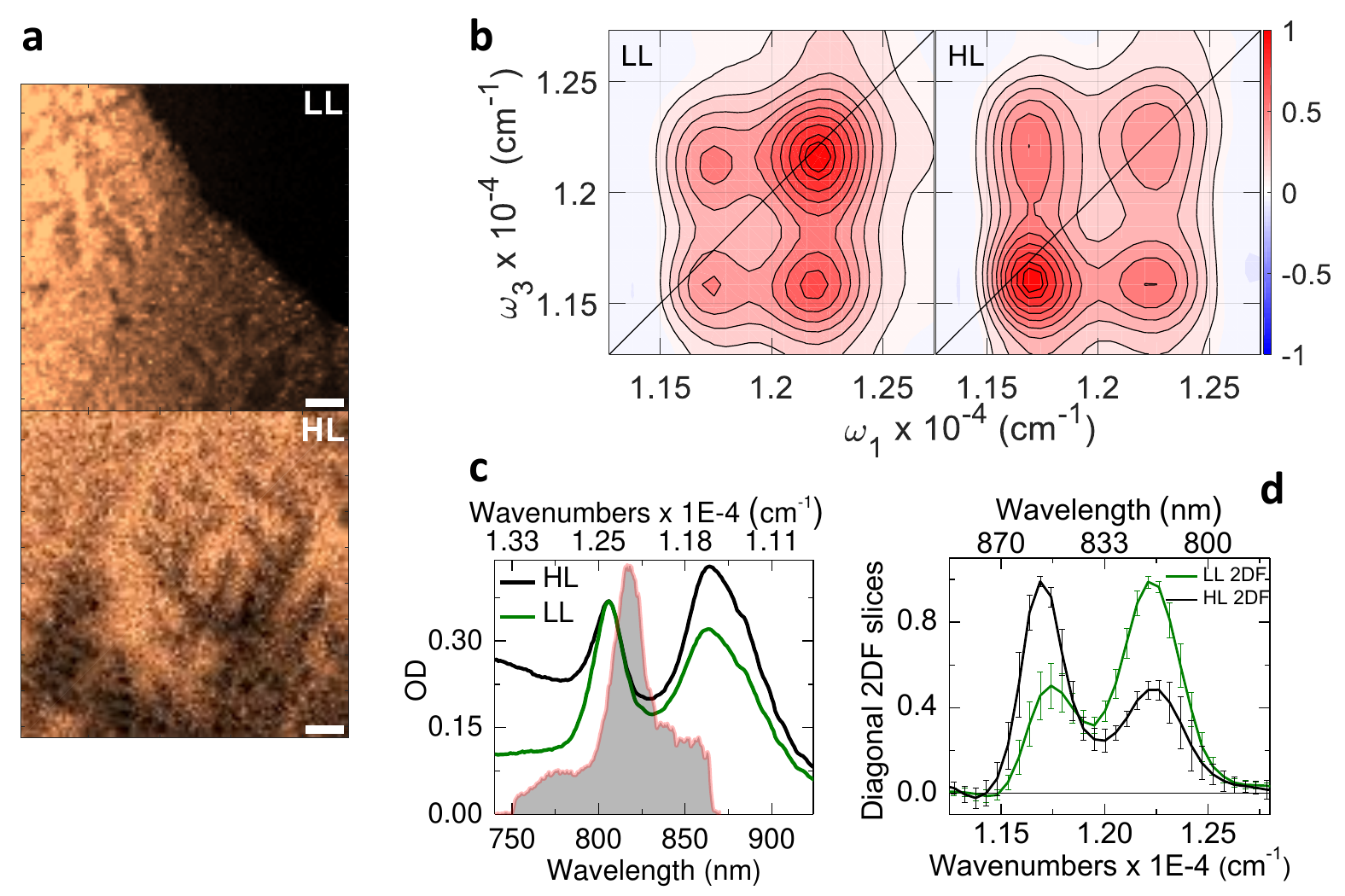}
	\caption{ SF--2DES on unmixed cells of \emph{Rps. palustris} grown under HL and LL light intensity conditions. \textbf{(a)} Confocal fluorescence images of a drop-dried film of photosynthetic bacteria \emph{Rps. palustris}. The bacteria were grown under LL (top panel), and HL (bottom panel) conditions. A 10 $\mu$m scale bar is shown for reference. The OD of the live cell solution from which the samples were prepared is shown in panel \textbf{c}, and the drop volume was measured to be $\sim$0.08 $\mu$lt through gravimetric analysis (SI). \textbf{(b)} Normalized absorptive 2D spectra at $t_2$ = 0 fs obtained by averaging 2D spectra from 5 different XY locations on the LL and HL fluorescence images, as shown in the left and right panels respectively. The spectra show well-resolved cross-peaks at $t_2$ = 0 fs. Contours are drawn at 10-90 $\%$ in steps of 10$\%$, 95$\%$ and 100$\%$. The frequency $\omega$ corresponding to the axes labels on the 2D plots corresponds to $\omega = \mid\omega'\mid/2\pi c$, where $\omega'$ is in $rad/fs$.\textbf{(c)} Linear absorption spectrum for LL and HL grown \emph{Rps. palustris} overlaid with the laser spectrum (grey area). The OD for the LL sample is scaled by a factor of 0.37 such both samples have the same OD at the B800 band. \textbf{(d)} Slices through the maxima of the upper and lower diagonal peaks of the LL and HL 2D spectra shown in panel \textbf{b}. The error bars are obtained from averaging LL and HL 2D spectra at 5 different locations on the fluorescence images of panel \textbf{a}. The solid black line across the 2D plot corresponds to the diagonal. All measurements were conducted at 300 K.
	}
	\label{fig:fig2}
\end{figure*}

Fig. \ref{fig:fig2} shows \emph{in vivo} measurements illustrating how the growth condition dependent perturbation of the electronic structure of purple bacteria manifests in the SF--2DES spectra. Fig. \ref{fig:fig2}a shows fluorescence images collected from a bacterial colony in unmixed samples of LL (upper panel) and HL (lower panel) bacteria. Multiple XY locations on these fluorescence maps were chosen to collect fluorescence detected 2DES spectra. Fig. \ref{fig:fig2}b shows the averaged absorptive fluorescence detected 2DES spectra for LL and HL bacteria. The positions of the two diagonal peaks correspond to the B800 (upper diagonal) and B850 (lower diagonal) excitonic manifolds of the LH2 antennae within the bacterial cells. The effect of growth conditions on spectral properties is reflected by the varying strength of the lower (B850) diagonal peak. This is better seen in Fig. \ref{fig:fig2}d, which compares slices through the maxima of the diagonal peaks for the LL and HL cases. The measurements show that between HL and LL conditions, the 2D diagonal peak strengths change by a factor of four, that is, the ratio of B850/B800 diagonal peak changes from 2:1 to 1:2. The measured changes are well above the error bars of the measurement, emphasizing that the SF--2DES spectrometer is able to distinguish between the two kinds of cells with a respectable signal-to-noise ratio. Higher error bars near the peak slopes in Fig. \ref{fig:fig2}d reflect the fact that any index shifts in the 2D spectra, resulting from trial-to-trial variations, are averaged over. Averaging the 2D spectra would also lead to broader than ideal 2D peakshapes. The strength of a 2D peak is a product of the absorption and emission transition strengths \cite{JonasARPC}, such that it depends on $\sim\mu^4$, where $\mu$ is the electronic transition dipole. Thus, a factor of 4 change on the diagonal peaks from HL and LL cells suggests that the absorption strength ($\mu^2$) for the B850 manifold decreases by a factor of 2 under low-light conditions. The cross-peak strengths are not perturbed appreciably because they depend on the product of B800 and B850 absorption strengths. The 2D cross-peaks also highlight the fact that despite the strong perturbation of the site energies of the \emph{BChl a} pigments on the B800 and B850 rings, the \emph{BChl a} transition dipoles at B800 and B850 sites remain coupled through Coulomb interactions. Fig. \ref{fig:fig2}c compares the absorption spectrum of HL and LL grown cells. It is seen that the spectral changes highlighted by the absorption spectrum, which reflect the strength $\mu^2$ of a given electronic transition, are less than that indicated by the 2D peaks because, as mentioned above, the 2D peak strengths depend on $\mu^4$, and thus are more sensitive to changes in the excitonic structure. Both LL and HL absorption spectra show a shoulder around 875 nm (B875 band) which corresponds to the absorption of the LH1 complex. The laser spectrum (shown as solid grey area) overlaps dominantly with the blue part of the B850 band, and excitation of the B875 band is minimal.
\FloatBarrier

\subsection*{SF--2DES on mixed samples}

The measurements discussed above establish that  spectral differences in the LL and HL cells can be clearly differentiated by a SF--2DES experiment. In order to demonstrate the spatial-resolution and ability to characterize complex samples provided by the SF--2DES spectrometer, we performed measurements on bacterial colonies with a spatially heterogeneous composition of the LL and HL cells. The idea behind these measurements was to simulate what one might expect in a heterogeneous thin film sample, such as those known in mixed hallide perovskites \cite{Huang2015,Kukura2016,Ginsberg2017}, and polymer-fullerene blends \cite{Asbury2014}. In a mixed sample, the overall 2D signal from a given point in space will be averaged over the HL and LL constituent cells. In order to deduce the ratio of HL:LL cells contributing to the 2D signal, the averaged HL and LL spectra in Fig. \ref{fig:fig2}b are chosen as the 2D basis spectra, and a linear least squares fit of the mixed 2D signal is performed. Such an analysis assumes that the HL and LL cells are equally fluorescent. Thus a given HL:LL ratio indicates the ratio of FWM signal from HL versus LL cells, rather than the ratio of the number of HL to LL cells. 

\begin{figure*}
	\centering
	\includegraphics[width=6.3 in]{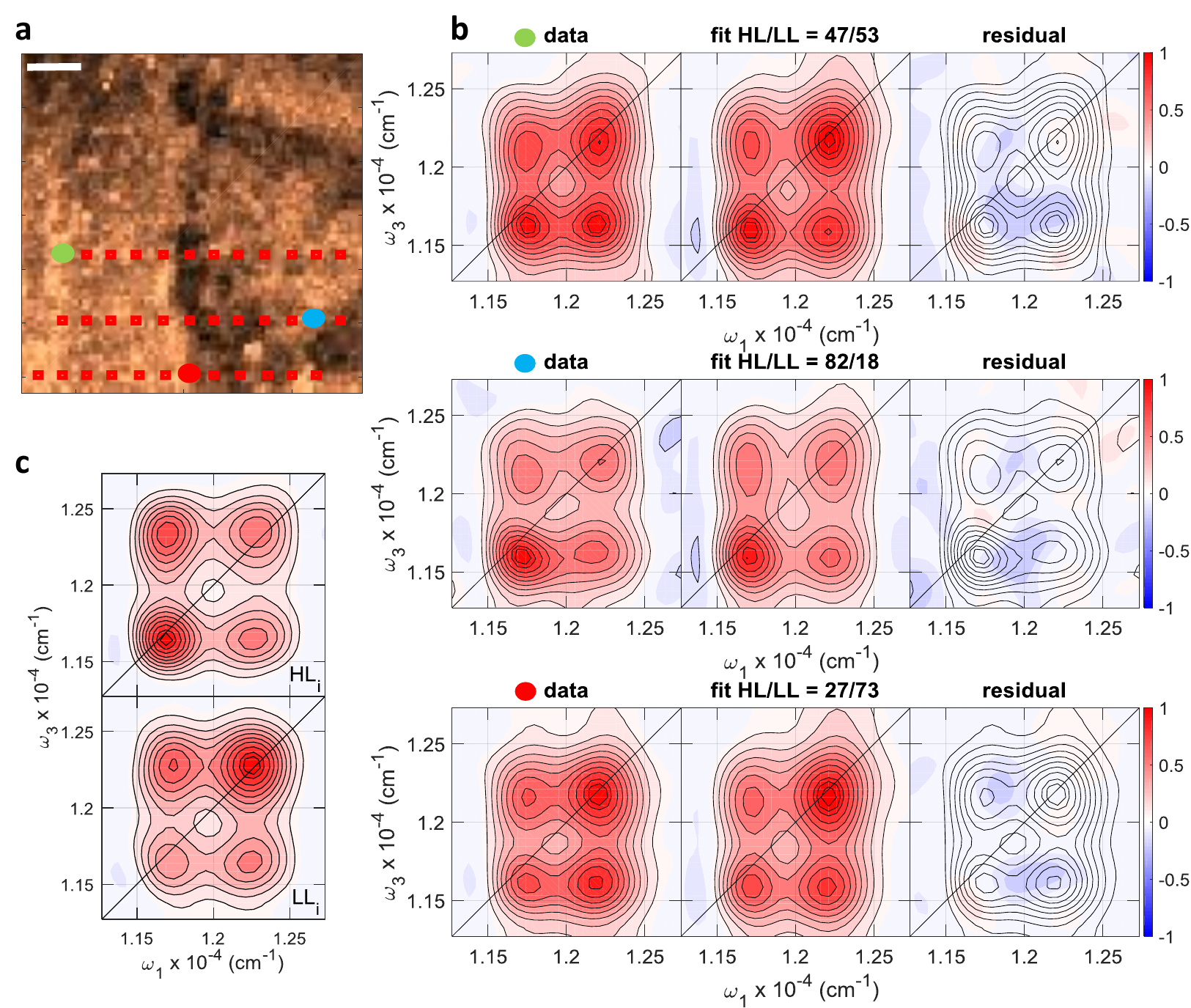}
	\caption{\textbf{(a)} Confocal fluorescence image from a mixed, LL and HL, drop-dried film of photosynthetic bacteria \emph{Rps. palustris}. Red squares correspond to all the locations where a 2D spectrum was collected, and is shown in the SI. The red, blue and green solid dots indicate locations for which the corresponding 2D spectra are shown in panel \textbf{b}. A 10 $\mu$m scale bar is shown for reference. \textbf{(b)} Normalized $t_2$ = 0 fs absorptive 2D spectra collected at the three solid dot locations shown on the fluorescence image in panel \textbf{a}. The three horizontal frames show the measured 2D spectrum (left frame), a linear least squares fit obtained by using the LL and HL 2D spectra (in Fig. \ref{fig:fig2}b) as the basis spectra to describe the spectrum collected from a mixed sample (middle frame), and the resulting residual (right frame). All the frames in a panel are normalized to the left frame. The middle frame also displays the ratio of purely HL and LL 2D basis spectra (shown in Fig. \ref{fig:fig2}b), which best fits the measured spectrum. \textbf{(c)} Absorptive $t_2$ = 0 fs 2D spectra for the HL and LL samples obtained by recirculating a live cell solution through a 200 $\mu$m pathlength sample cell using a peristaltic pump. These spectra represent the ideal scenario (represented as LL$_i$, HL$_i$), where there is no measurable photobleaching. The OD of the sample in 1 mm pathlength corresponds to that shown in Fig. \ref{fig:fig2}c. Contours are drawn at 10-90 $\%$ in steps of 10$\%$, 95$\%$ and 100$\%$. The frequency $\omega$ corresponding to the axes on the 2D plots corresponds to $\omega = \mid\omega'\mid/2\pi c$, where $\omega’$ is in $rad/fs$. The solid black line across the 2D plot corresponds to the diagonal. All measurements were conducted at 300 K.
	}
	\label{fig:fig3}
\end{figure*}

Fig. \ref{fig:fig3}a shows the confocal fluorescence image obtained from a bacterial colony which has spatially heterogeneous composition of HL and LL cells. A 2D spectrum was collected at several points on this image, marked in red squares. The points are separated horizontally by $\sim$5 $\mu$m. Three such 2D spectra, from the locations corresponding to the red, blue and green solid dots, are shown in Fig. \ref{fig:fig3}b. A fit of the 2D spectrum from each location as a linear combination of the 2D basis spectra is shown in the middle frame of Fig. \ref{fig:fig3}b, and the residual is shown in the right frame. The plots for all other locations, and the fitting details, are provided in the SI. The three chosen locations approximately show the extremes of the expected spatial variations, that is, a dominantly HL (middle) or LL (bottom) spectrum, as well as a spectrum which is an equal mixture (top). The fits and the residuals are shown on the same scale as the data, and show average fit errors of $\sim$10$\%$, across all the locations marked in red squares in Fig. \ref{fig:fig3}a, emphasizing the capability of SF--2DES to resolve the level couplings manifested in 2D peakshapes and their strengths, into their heterogeneous constituents, across space. Morphological variations in the pump-probe signal are already known for a number of systems \cite{Huang2015, Kukura2016, Warren2011, Asbury2014}. However, in terms of resolving such variations into 2D peakshapes so as to infer couplings between electronic transitions, the capability provided by SF--2DES is unprecedented. 

Fig. \ref{fig:fig3}c shows the LL and HL 2D spectra from unmixed sample solutions of live cells which were recirculated using a peristaltic pump. These spectra are considered `ideal cases' (LL$_i$ and HL$_i$) because they exhibit no photobleaching artifacts. Changes in the sample OD and FWM signal before and after these experiments were not measurable beyond trial to trial variations. Normally, Mie scattering encountered in flowing similar colloidal solutions \cite{Dahlberg2017, Zigmantas2016} becomes a dominant source of noise in interferometric detection. However, despite the all collinear geometry, the combination of spectrally filtering the fluorescence from the laser, and high-frequency lock-in detection, allows us to make \emph{in vivo} measurements with relative ease. A comparison of LL$_i$ and HL$_i$ spectra to the basis 2D spectra in Fig. \ref{fig:fig2}b, and with the 2D spectra on mixed samples (Fig. \ref{fig:fig3}b), shows vertical distortions in the spectra collected from immobilized samples. This distortion is partly caused by FWM signal degradation due to photobleaching of \emph{BChl a} pigments within the cells (see Discussion).  
\FloatBarrier
\subsection*{SF--2DES spatial resolution}

A diffraction-limited optical imaging system images an ideal point as a three-dimensional light intensity distribution also called the point spread function (PSF) \cite{MertzBook}. For a conventional fluorescence imaging system, the PSF is defined as $H_{id,conv}$ = $I(r,z)$, where $I(r,z)$ denotes the excitation light intensity distribution at the focus, and the subscript $id$ denotes that it is an ideal diffraction-limited PSF. Here $r$ and $z$ are lateral and axial coordinates derived from optical units $v$ and $u$, respectively \cite{Wilson2011}.
For confocal microscopy, the PSF depends linearly on the excitation intensity, as well as on the size of a point detector, and is given by $H_{id,conf}(r,z) = I(r,z)[I(r,z)\otimes D(r)]$. The second term results from a two-dimensional convolution of the excitation intensity with a detector $D(r)$, such as a pinhole, and is responsible for optical sectioning in confocal microscopy. When the pinhole is large compared to the size of the Airy disc, the second term becomes constant, such that $H_{id,conf}$ is approximately the same as conventional PSF $H_{id,conv}$, that is, $H_{id,conf}(r,z) \sim H_{id,conv}$ \cite{Wilson2011}. This is also true for the confocal imaging part of the current experiment because a large detection pinhole (200 $\mu$m pinhole diameter compared to a sub-micron Airy disc diameter) renders the effect of a detection pinhole on the confocal PSF negligible. 

Assuming a large detection pinhole, for the case of two-photon (TP) microscopy, the excitation point spread function (PSF) depends quadratically on the excitation intensity, that is, $H_{id,TP}(r,z) \sim H^2_{id,conf}(r/2,z/2) = I^2(r/2,z/2)$ \cite{Sheppard1995}. The functional dependence of intensity on $r,z$ is different in the TP case due to the difference of $1/2$ in the excitation wavelength. In analogy, the FWM signal in a fluorescence-detected 2DES experiment is generated by one light-matter interaction of the sample with each pulse in sequence of pump and probe pulse pairs, and therefore depends linearly on the pump and probe intensities. Thus, the volume of the FWM excitation PSF, which dictates the volume of the sample contributing to the FWM signal, will be a product of pump and probe intensities, that is, $H_{id,FWM}(r,z) \sim I^2(r,z)$. Note that unlike in the case of TP microscopy, the FWM process and fluorescence detection happens at approximately similar wavelengths, and therefore, $r,z$ do not have a factor of $1/2$. Consequently, the spatial resolution dictated $H_{id,FWM}$ will be better than that provided by TP microscopy. 

For the water objective (NA 1.2) used for SF--2DES experiments at a peak laser wavelength of $\sim$ 820 nm, the ideal FWM PSF $H_{id,FWM}(r,z)$ is shown in Fig. \ref{fig:fig4}, left panel. The lateral and axial FWHMs from Fig. \ref{fig:fig4}, left panel are $\sim$ 0.25 $\mu$m and $\sim$ 0.69 $\mu$m, respectively. These represent the diffraction-limited spatial resolution obtainable from a  SF--2DES spectrometer. As with confocal TP microscopy \cite{Sheppard1995}, the use of a smaller pinhole could further improve the spatial resolution of SF-2DES.

\subsection*{Estimation of SF--2DES sensitivity}

In order to estimate the sensitivity of SF-2DES compared to conventional heterodyne-detected 2DES studies, we perform an order of magnitude estimation of the number of cells contributing to the FWM signal measurements on bacterial colonies, and compare to recent \emph{in vivo} studies \cite{Dahlberg2017,Zigmantas2016} on cells using conventional 2DES. 
  
\begin{figure}[h!]
	\centering
	\includegraphics[width=3.2 in]{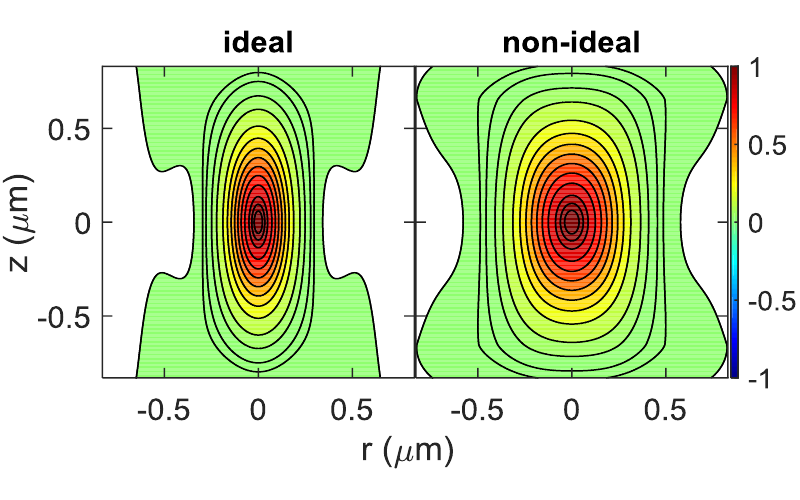}
	\caption{\textbf{(left panel)} The expected ideal diffraction-limited FWM PSF for the case when the pump-probe laser beams with 820 nm peak laser wavelengths, are focused to a point in the sample using a water objective of NA 1.2, and a peak laser wavelength of 820 nm. \textbf{(right panel)} The non-ideal FWM PSF calculated from the ideal FWM PSF by convoluting it with a 0.54 $\mu$m diameter bead to approximate the measured experimental broadening. Contours are drawn at 0.1$\%$, 1$\%$, 2$\%$, 5$\%$, 10-90$\%$ in steps of 10$\%$, 95$\%$ and 98$\%$.
	}
	\label{fig:fig4}
\end{figure}

A confocal fluorescence image of a 0.5 $\mu$m bead measured with the imaging part of the SF--2DES spectrometer shows an average FWHM lateral diameter of 0.83 $\mu$m. The image and the fitting details are provided in the SI. However, the expected FWHM, obtained by convoluting the diffraction-limited lateral confocal PSF with a 0.5 $\mu$m bead, is $\sim$ 0.52 $\mu$m (SI). This implies that the experimentally broadened lateral confocal PSF deviates from the ideal case by a factor of $\sim$ 0.83/0.52 = 1.6. Based on this deviation, in order to estimate the experimental non-ideal FWM PSF, we first convolute the ideal confocal PSF $H_{id,conf}$, with a bead of a given diameter, such that the resulting confocal PSF has a lateral width of 1.6x the ideal lateral width. A bead diameter of 0.54 $\mu$m approximates this experimental broadening, and the resulting non-ideal confocal PSF (shown in the SI) is denoted as $H_{nid,conf}$, where the subscript $nid$ denotes the non-ideal case. The non-ideal FWM PSF $H_{nid,FWM}$ is then calculated as $H_{nid,FWM} \sim H^2_{nid,conf}$. Fig. \ref{fig:fig4}b shows the estimated $H_{nid,FWM}$ which dictates the experimental spatial resolution. As discussed above, the FWM PSF in Fig. \ref{fig:fig4} (right panel) approximates the experimental lateral broadening of the PSF measured through the confocal image (SI). We note that the above analysis assumes that the experimental broadening of the diffraction-limited PSF along the \emph{axial} direction is also affected in the same way as measured along the lateral direction, which is a reasonable assumption for an order of magnitude estimation.

The volume enclosed by the 10$\%$ contour in the FWM PSF in Fig. \ref{fig:fig4}, defines the region inside which the excitation probability of a FWM signal is >10$\%$ of the maximum excitation probability in the experiment, that is, the sample volume from which >90$\%$ of the FWM signal contributes. For the ideal case in the left panel, the enclosed 90$\%$ volume is $\sim$ 0.19 $\mu$m$^3$, and increases to $\sim$ 0.55 $\mu$m$^3$ (0.55 femtoliters) when the FWM PSF is non-ideal (right panel). In comparison, in a conventional 2DES experiment with a 200 $\mu$m pathlength and a 100 $\mu$m diameter spot size, the pump-probe overlap volume is $\sim$1.6 x 10$^6$ $\mu$m$^3$, which is over six orders of magnitude larger than in our SF-2DES setup. Assuming the same starting OD for the experiments, we estimate $\sim$25 cells/$\mu$m$^3$ of the solution. This gives an upper estimate of $\sim$ 10$^3$ cells in a dried drop of ~0.08 $\mu$L initial volume contributing to the FWM signal in SF--2DES (SI), versus 10$^7$ cells solution for the case of 2DES experiments. Thus, fluorescence-detection allows measurements with a respectable signal-to-noise ratio, with nearly 4 orders of magnitude fewer cells.

\section*{Discussion}
A reduction in the probed volume by approximately six orders of magnitude of SF-2DES compared to conventional 2DES is due to the optical sectioning enabled by the FWM PSF of the microscope. As discussed above, the lateral and axial resolution provided by non-linear imaging through SF--2DES will be better than two-photon imaging, and could be further enhanced with a smaller pinhole for confocal detection \cite{Sheppard1995,Wilson1995}. Further  resolution enhancement could be possible with the use of metallic nanostructures, which have served as optical antennas \cite{vanHulst2011} that allow sub-diffraction-limited focussing of electromagnetic fields. Optical antennas have allowed near-field imaging with femtosecond lasers with spatial resolution down to a few tens of nanometers \cite{Raschke2016,Potma2014}. Plasmonic enhancement also allows detection of weak FWM signals, even reaching the single molecule limit \cite{Potma2014}. Similar resolution of a few tens of nanometers has also been demonstrated in variation of stimulated emission depletion microscopy (STED) with picosecond time resolution \cite{Ginsberg2015}. The collinear geometry, facile manipulation of the spectra of individual pulses, and frequency filtering by the lock-in amplifier makes it possible to integrate the SF-2DES spectrometer with the above super-resolution imaging approaches. Moreover, the collinear geometry reduces distortions of the 2D peak shapes due to directional filtering and phase-matching \cite{Jonas2007}.  For applications involving optically thick samples, such as deep tissue imaging, resonant fluorescence re-absorption, which is not an issue in TP microscopy, could cause higher order non-linear effects such as sequential and parallel signal cascades \cite{Albrecht1998,Fleming1999}, or absorptive/dispersive signal distortions \cite{Jonas2007}. 

A comparison of Figs. \ref{fig:fig2}c and \ref{fig:fig3}b with Fig. \ref{fig:fig3}c shows that the ideal (LL$_i$ and HL$_i$) 2D spectra are distorted vertically when measured on immobilized samples. This vertical broadening is in part due to averaging the spectra over multiple locations on the fluorescence image, but a majority can be accounted for by FWM signal degradation due to photobleaching. Experimentally, the $t_1$ time delay is scanned for each successive value of $t_3$. Thus, a $t_3$ slice (corresponding to $t_1$ = 0 fs) is acquired over a longer time period during which photobleaching can accumulate, leading to the vertical distortion along $\omega_3$. Assuming an exponential signal degradation with each successive time index in the experiment, a simulation of this effect (shown in Figure S1 of the SI) captures the observed 2D peak broadening due to photobleaching. The simulations show that photobleaching as much as 80$\%$ of the signal during the data collection time of 45 seconds causes 2D peaks to merge together, with cross-peaks appearing as shoulders. Based on the well-resolved peaks seen in Fig. \ref{fig:fig3}b, we estimate an upper limit of 60$\%$ signal degradation during the data collection time. Signal degradation is expected to effect dynamics along the pump-probe waiting time $t_2$, however the couplings probed by the distorted 2D peakshapes at $t_2$ = 0 fs, remain surprisingly robust, as indicated by the measured well-resolved cross-peaks. Improvements such as larger time steps made possible by physical undersampling of the signal, that is, signal detection at $\omega_{eg}-\omega_{R1(2)}$ frequency, can easily reduce the experiment time by order $N^2$, where $N$ is the total number of time steps in the experiment. For example, in the current experiment, using $t{_1}$ and $t{_3}$ time steps of 10 fs instead of 5 fs, reduces the data collection time to $\sim$12 seconds without any frequency aliasing effects (see SI). Reduced dimensionality experiments which extract particular signals of interest along the waiting time \cite{Seckin2015}, can also lead to substantially faster data collection and improved signal-to-noise ratios.  Additionally, greater sample preparation efforts to reduce photobleaching, by using oxygen scavengers or low temperature could reduce spectral distortions due to photobleaching. Lower laser repetition rates, down to $\sim$ 1 MHz, are also known to reduce photobleaching and increased fluorescence signal levels \cite{Hell2006}. We note that although we observe photobleaching effects, our phase-modulation-based SF-2DES approach should be considerably less susceptible to distortions from photobleaching than phase-cycling methods. Since the phase-cycling approach relies on the acquisition of 27 consecutive scans that are combined to extract the 2D spectrum \cite{Brixner2018}, to achieve a high signal-to-noise ratio it is critical that the signal level be stable throughout all 27 scans to enable effective signal isolation.  

The background-free nature of fluorescence-detection, coupled with the development of high quantum efficiency single-photon detectors has made  single molecule fluorescence detection routine \cite{MoernerReview}. Thus, in terms of scalability of 2DES to very low numbers of absorbers,  fluorescence-detection approach may hold more promise than heterodyne-detection of a radiated electric field, as typically employed in ensemble 2DES experiments. Our experiments demonstrate that F-2DES is highly sensitive: for experiments in which the sample was recirculated (Fig. \ref{fig:fig3}c), a respectable signal-to-noise ratio was attainable with ODs $\sim$ 5x lower (in a 200 $\mu$m pathlength) than those used in previous $\emph{in vivo}$ studies \cite{Dahlberg2017, Zigmantas2016}, with the estimated probed volume $\sim$ 6 orders of magnitude smaller due to the FWM PSF created by the microscope objective.  Varying energy transfer pathways have been reported \cite{vanHulst2013} for isolated LH2 complexes through single-photon counting. SF--2DES offers a promising route towards observing such effects using 2D spectroscopy, and resolve the debate \cite{Scholes2017} surrounding the timescale of the survival of coherent wavepackets through sub-ensemble resolution. For non-fluorescent samples, the SF--2DES spectrometer can be modified for 2D photocurrent detection \cite{Cundiff2013}, or for heterodyne detection by delaying the 4$^{th}$ pulse and routing it around the sample as the local oscillator \cite{Martin2016}. Domain heterogeneity in polymer-fuller blends, or thin films of perovskites and TIPS-pentacene, could also quench the photoluminescence at certain locations in the film \cite{Ma2016}, making the above modification necessary in order to probe the domain-dependent carrier delocalization dynamics in these materials that has been observed in pump-probe microscopy studies \cite{Asbury2014,Huang2015,Wong2015}.

In conclusion, we have demonstrated \emph{in vivo} spatially-resolved measurements of 2D electronic spectra from colonies of photosynthetic bacteria grown under different light intensity conditions and have resolved the resulting differences in excitonic structure of the antenna proteins inside the cells. The Coulomb couplings between electronic levels and their perturbations with growth conditions are reflected as spatial variations in the well-resolved 2D peak amplitudes that arise from spatial heterogeneity in the bacterial colonies. The fluorescence-detection approach we adopt offers significant sensitivity improvements over conventional heterodyne detection. Employing phase-modulation rather than phase-cycling enables imaging at high repetition rates, and produces spectra with a high signal-to-noise ratio without the need to combine multiple phase-cycled scans that could be susceptible to photobleaching. This work serves as a proof-of-concept demonstration of the broad applicability of the SF--2DES approach towards resolving the connections between morphological variations, the resulting electronic couplings, and ultimately the performance of a variety of light-harvesting materials.

\begin{footnotesize}
\section*{Methods}
\subsection*{SF--2DES setup}         
Here we provide a concise description of SF--2DES spectrometer, shown in Fig. \ref{fig:fig1}, and refer the reader to the SI for additional details. The experimental layout is based on the original fluorescence-detected 2DES design by Marcus and co-workers \cite{Tekavec2007}. The output from a commercial broadband 83 MHz Ti:Sapphire oscillator (Venteon One) is sent into an SLM-based pulse-shaper (MIIPS 640P, Biophotonic Solutions) for dispersion pre-compensation. Laser pulses from the pulse shaper of $\sim$20 fs FWHM pulse duration are split by a 50:50 beamsplitter and routed into two Mach-Zehnder (MZ) interferometers. Within each of the two MZs, the pulse is further split using 50:50 beamsplitters (BS), and each arm is tagged with a unique phase that is modulated at radio frequency using acousto-optic modulators (AOMs). The phase modulation frequencies for the four arms are denoted by $\Omega_{1-4}$, such that when the split pulses are recombined at BS3 and BS5, within MZ1 and MZ2 respectively, the pulse amplitude modulates at the difference frequencies of the AOMs, that is, $\Omega_{12}$ and $\Omega_{34}$  for MZ1 and MZ2, respectively. One output port from BS3 and BS5 is used to generate reference frequencies REF1 and REF2 for MZ1 and MZ2 respectively, which modulate respectively at $\Omega_{12}$ and $\Omega_{34}$ difference frequencies. These frequencies are used for phase-sensitive lock-in detection. Beams from the other two output ports (one from each of the two MZs) are recombined at BS6 to produce a train of  four collinear pulses. The relative pulse time delays are controlled by translational stages resulting in an all-collinear train of 4 pulses separated by time delays $t_1$ between the first two pulses, $t_2$ between the 2$^{nd}$ and 3$^{rd}$ pulses, and $t_3$ between pulses 3 and 4  in the sequence. 

The train of 4 phase-modulated pulses passes through a OD4 875 SP filter (Edmund Optics), and is routed into a confocal microscope (Olympus 1X51). The pulse train is transmitted through a 875 nm dichroic mirror (Semrock) inside the microscope before reaching the water objective (Olympus PlanApo 60x, NA1.2) with the objective collar set to 0.17. The immobilized sample is mounted on an XY scanning piezo stage (Piezo Instruments, P-612.2SL). The fluorescence is collected in the epi-detection geometry and directed by a DCM to be spatially filtered through a 200 $\mu$m detection pinhole, followed by optical filtering through an OD 4 875LP filter. For linear fluorescence imaging experiments, the filtered fluorescence is focused onto a single photon counting APD1 (not shown in Fig. \ref{fig:fig1}). The photon counts registered by APD1 are counted using a counting circuit (not shown in Fig. \ref{fig:fig1}) for each XY position of the computer controlled piezo controller. 

For SF--2DES measurements on the imaged sample, or for fluorescence detected 2DES experiments in which the sample is circulated, the filtered fluorescence is routed by a flip mirror (FM) and focused on another APD operating in linear mode (not shown in Fig. \ref{fig:fig1}). $t_1$ and $t_3$ delays are scanned from 0 to 90 fs in steps of 5 fs, and t$_2$ delay is scannable from 0 to 800 ps, but is fixed at $t_2$ = 0 fs for the experiment. The four-wave mixing signal generated by the sample oscillates at the difference frequencies $(\Omega_3 - \Omega_4) \pm (\Omega_1 - \Omega_2)$, where the (positive) negative sign corresponds to (non-rephasing) rephasing 2D signals. The oscillating signal is demodulated using a lock-in amplifier (Zurich Instruments, HF2LI). Physical undersampling of the oscillating signal, as well as phase-sensitive detection, is achieved by signal detection relative to the reference signals generated using REF1 and REF2 MZ ports. The reference signals are connected to the lock-in amplifier parallel reference channels corresponding to rephasing and non-rephasing signals. The references for all the SF-2DES experiments are set at 826 nm. The AOM frequencies are set at $\Omega_1$ = 80.111 MHz, $\Omega_2$ = 80.101 MHz, $\Omega_3$ = 80.029 MHz and $\Omega_4$ = 80.0 MHz through a common clock, such that the resulting signals oscillate at 19 kHz (rephasing) and 29 kHz (non-rephasing). The data collection time for each 2D spectrum was $\sim$45 secs for 5 fs $t_{1,3}$ steps. The data collected for each $t_2$ delay, is Fourier transformed with respect to $t_1$ and $t_3$, resulting in a correlated map of absorption ($\omega_{1}$) and detection ($\omega_{3}$) frequency axes as shown in Fig. \ref{fig:fig1}. A flip mirror (FM) is used to switch between the imaging and spectroscopy modalities. A 2D spectrum can thus be generated for all the desired points on the fluorescence image.

 For experiments where the sample is circulated, a 200 $\mu$m pathlength sample cuvette and an air objective (Olympus LUCPlanFLN 40x, NA0.6) are used for the measurements. The sample is circulated through the cuvette using a peristaltic pump at flow rates of ~190 ml/min. For linear fluorescence imaging, the total incident power on the sample is $\sim$0.125 $\mu$W/pulse, corresponding to a fluence of 0.26 $\mu$J/cm$^2$/pulse. The binning time at each piezo location is 5 ms. For SF--2DES experiments on immobilized and circulating samples, the spectra were collected with a total power of 7.5$\mu$W/pulse, corresponding to a total fluence of 12-16 $\mu$J/cm$^2$/pulse. At 83 MHz repetition rate, the total pulse energies for these fluences is $\sim$0.09 pJ/pulse, such that the excitation probability at the center of the pulse is less than 0.1$\%$. Low excitation probability is chosen so as to minimize multiple excitations on the molecule by the same pulse.

\subsection*{Sample preparation and handling}
Cells of \emph{Rps. palustris} strain 2.1.6 were grown anaerobically in closed, flat-sided bottles using C-succinate media at 30 $^o$C under anaerobic conditions at different light intensities. High light (HL) cultures were placed between two 100W incandescent bulbs to give either high-light (HL, $\sim$150 $\mu$mol photons s$^{-1}$ m$^{-2}$) and lower-light (LL, $\sim$100 $\mu$mol photons s$^{-1}$ m$^{-2}$). The cells were harvested when fully adapted, washed in 20 mM MES, 100mM KCl, pH 6.8, aliquoted, flash frozen and stored at -20 $^o$C until required. For the experiments, a thick aliquot of cells was suspended in the buffer such that the final OD in a 1 mm cuvette was 0.38 for HL and 1.03 for LL cells. Using the microscope eyeppiece, it was checked that the cells in a drop of a diluted sample solution were motile. No oxygen scavenging agents were used in the cell solutions. For experiments with circulating sample, approximately 10 ml of starting cell solution was circulated through a 200 $\mu$m cuvette. During the experiment, the sample was kept in an ice-cooled water bath using a Peltier cooler. For experiments on immobilized samples, a ~0.08 $\mu$L drop of the cell solution was dispensed on a 0.17 mm coverslip using a syringe tip (BD U-100 31G), and the drop was allowed to dry under nitrogen gas. The average volume of the syringe-dispensed drop was measured to be 0.079 $\pm$ 0.002 $\mu$L using gravimetric analysis (by measuring the average weight of drop of 15 syringe-dispensed drops in three trials). The coverslips were cleaned by leaving them in a bath of 6M HCl in ethanol on a hot plate at 100 ${^o}$C for $\sim$2 hours, followed by sonication for a few hours in a bath of distilled water and ethanol. However, for the experiments on mixed HL and LL samples, the cover slips were only cleaned with distilled water such that the surface of the cover slips was hydrophobic enough to not allow the HL and LL drop to spread and mix together uncontrollably. For the mixed HL/LL experiments, initially a HL drop was allowed to dry followed by dispensing a LL drop close to the original drop. The LL drop was manually dragged (using a syringe tip) near the dried HL drop until it partially mixed with the HL drop. The final drop was allowed to dry under nitrogen, and the coverslip was sealed against a glass slide using double-sided tape on the inner surface and Scotch tape on the outer surface. The diameter of the final drop after drying was measured to be $\sim$1 mm. Typically, the experiments were conducted within $\sim$1 hour of sample preparation. Using the microscope eyepiece it was observed that after more than 3 weeks of drying the cells under Nitrogen, the cells regained motility once a drop of water was poured on the dried drop.

For fluorescence imaging in order to estimate the experimental confocal PSF, commercially available fluorescent polystyrene beads (Fluoresbrite 763 Carboxylate Microspheres 0.50 $\mu$m) were diluted by a factor of 10$^8$ in distilled water, and a 20 $\mu$L drop was drop-dried on a 0.17 mm microscope coverslip. The absorption and emission wavelength of the dye inside the beads was 763 nm and 820 nm respectively. The laser spectrum used for imaging the bead was filtered with a 800 shortpass excitation filter (Chroma), and the fluorescence was detected with a 815 longpass detection filter (Chroma). The peak laser excitation wavelength was 780 nm. 

\section*{Acknowledgements}

The authors acknowledge the AFOSR Biophysics program for support of this research under grant FA9550-15-1-0210. R.J.C. and A.T.G. gratefully acknowledge support from the Photosynthetic Antenna Research Center, an Energy Frontier Research Center funded by the U.S. Department of Energy, Office of Science, Office of Basic Energy Sciences under Award Number DE-SC 0001035. The authors also thank Julia Widom, Andrew Marcus, and Eric Martin, Travis Autry of the Cundiff group for the enthusiastic help in building the fluorescence-based 2D setup, and Damon Hoff for advice regarding the microscope setup and sample preparation. 

\section*{Author Contributions}

J.P.O. conceived the idea for the method. V.T. and Y.A.M designed the experiment in consultation with J.P.O. R.J.C. and A.T.G. gave advice on the sample choice and preparation and provided the samples. V.T. and Y.A.M. performed the experiments and data analysis. V.T., Y.A.M. and J.P.O. wrote the paper with input from all of the authors. 

\end{footnotesize}



\end{document}